\documentclass[letterpaper]{article}
\usepackage{aaai}
\usepackage{array}
\usepackage{times}
\usepackage{helvet}
\usepackage{courier}
\usepackage{amssymb}
\usepackage{amsmath}
\usepackage{graphicx}
\usepackage{algorithm2e}
\usepackage{booktabs}

\nocopyright
\title{Planning for the Efficient Updating of Mutual Fund Portfolios}

\author{Tomás de la Rosa \\
J.P. Morgan, AI Research \\
Madrid, Spain \\
tomas.delarosaturbides@jpmorgan.com
}

\begin{document}
\maketitle

\begin{abstract}
Once there is a decision of rebalancing or updating a portfolio of funds, the process of changing the current portfolio to the target one, involves a set of transactions that are susceptible of being optimized. This is particularly relevant when managers have to handle the implications of different types of instruments.
In this work we present linear programming and heuristic search approaches that
produce plans for executing the update. The evaluation of our proposals shows cost improvements over the compared based strategy. The models can be easily extended to 
other realistic scenarios in which a holistic portfolio management is required
\end{abstract}

\maketitle

\section{Introduction}
\noindent In the context of individual managed accounts, portfolio updates are generally handled by back-office processes that execute the transactions needed to achieve the target allocation.   Regardless of the investment strategy \cite{li2014online,xidonas2020robust}, any recurrent allocation decision yields a target portfolio \cite{ben2000robust}, so a set of transactions are executed to change the current portfolio to the target one. For instance, a mean variance optimization approach \cite{jorion1992portfolio} will produce a new allocation for the next period, or a rule-based strategy might stick to a predefined allocation and trigger rebalancing \cite{zilbering2015best} when the current weights drift away from the strategic asset allocation.  
When portfolios hold funds or equivalent collective investment schemes, the update process needs to optimize the transactions to minimize costs or include additional investor non-pecuniary preferences.  The prevalence of robo-advisors \cite{phoon2017robo,jung2018robo} demonstrates the high levels of automation achieved for various rebalancing strategies.  However, a more holistic portfolio management approach, should consider that portfolio updates involve instruments of different types, denominated in different currencies or traded at different venues. Additionally, investors might have different accounts with different tax treatment or transaction restrictions.  Consequently, a simple strategy such as “sell allocations above target weights and buy allocations under target weights” is no longer appropriate.   We argue that academic focus on incorporating costs into portfolio selection \cite{das2013online} has overshadowed the planning of portfolio updates. An optimized plan for updating portfolios benefits both investors and back-office intermediaries involved in the transactions.  

A particular scenario of interest is the Spanish tax-deferral regime known as 'Traspasos', where individual investors in mutual funds can defer the capital gain taxes by using the proceeds from redeeming a mutual fund to subscribe shares of another fund. Rather than considering redemption and subscription as two trades, the joint operation is treated as a "switch" or a transfer without triggering a tax event.  However, this benefit does not apply universally to all collective investment schemes.  The most remarkable exception is Exchange-traded Funds (ETFs). Buying or selling ETFs in the market is considered as separate trades, therefore planning the update of the portfolio is more complex when it has a combination of transferable mutual funds and exchange-traded instruments.
Planning portfolio updates also gets complicated when the target allocation is decided for investors with multiple accounts \cite{idzorek2023personalized}, holding different types of instruments and have diverse tax treatments.

In this work we focus on developing solutions for the "Traspasos" use case, and then leave the discussion to analyze how our approach applies to other scenarios.
The main contributions of this work are twofold: (1) formalizing the portfolio update as a combinatorial problem, and (2) proposing the use of heuristic search to find the optimal update plan. The paper is structured as follows: Firstly, we describe the base case of planning fund switches for a portfolio holding transferable mutual funds. Secondly, we introduce the fund multi-type case, which presents a combinatorial problem due to the various transaction alternatives. We then model the problem as a search task, employing classical AI search algorithms for resolution. We evaluate the benefits of the approach in terms of transaction number and costs. Furthermore, we discuss potential extensions to inspire further research in this domain. Finally, we present the key insights we gained throughout the development of the project.

\section{Mutual Fund Switches}
\label{sec:singleswitch}
In the base case, we assume all portfolio holdings are mutual funds (i.e., they belong to the same type of instruments) and all update operations are transfers from one fund to another (i.e., single type of transaction). We call this case the \textit{fund switching problem}.

Before solving the problem we get the money flows from the difference between the target portfolio and the current portfolio. The inputs to the problem are outflows $p_{\rm out}$, inflows $p_{\rm in}$,  and cost function $c_{ij}$ 
that express the cost of switching 1 monetary units from fund $i$ to fund $j$. 

The fund switching problem can be easily modelled with linear programming (LP) as 
the balanced version of the so called Hitchcock–Koopmans transportation problem \cite{fulkerson1956hitchcock,tokuyama1995efficient}. In this transportation analogy, goods are the money value of funds (i.e., the decision variables), the supply sources correspond to the outflows and the demand destinations correspond to the inflows. The problem objective is to minimize the total costs of the switches (i.e. the transportation costs). The model is formally expressed in Eq~\ref{eq:lp_model}, where $x_{ij}$ is the amount transferred from $i$ to $j$.

\begin{equation}
    \label{eq:lp_model}
    Min\, Z= \sum_{i=0}^m \sum_{j=0}^n x_{ij} c_{ij}
\end{equation}
 \begin{align*}
    s.a. & \sum_{j=1}^n x_{ij} \leq p_{\rm out_i}  \\
         &\sum_{i=1}^m x_{ij} \geq p_{\rm in_j} \\   
         & x_{ij} \geq 0 \quad \quad \forall i j \\
 \end{align*}

\section{Multi-type Transaction Updates}
\label{sec:lp-dual}
In the general case, the LP transportation problem is no longer applicable if in addition to the switching transactions there are other ways
to move the money, and the order or timing of transactions is relevant. Let's consider an example of a managed account with 100k(€), which follows the model portfolio outlined in Table~\ref{tab:ssa}. The account is mandated to rebalance quarterly to the target weights. As the current allocations have deviated from the target, the indicated money outflows and inflows should be implemented to achieve the necessary rebalance update. In this case, exchange-traded securities (i.e., ETFs and ETC) can be bought or sold at any time during the day, but their allocated values cannot be transferred as a single operation as before. Instead, we must combine a buy transaction with a preceding sell transaction to achieve an equivalent outcome. Similarly, when allocating value from transferable funds to a non-transferable security, we must first redeem the shares and utilize the resulting proceeds to fulfill the new allocation.  The individual  subscription (buy) or redemption (sell) of mutual funds are transactions with typically end-of-day valuation. For the purpose of the model they are equivalent to ETFs market trades that increase or decrease the account cash balance. 

\begin{table*}[ht]
    \centering
    \begin{tabular}{clcccrr } 
       \toprule
       ID & Investment  & Target \% & Current \% & Transferable &  outflows (€) & inflows (€) \\ 
       \midrule
       MM & Money Market Fund & 10.0 & 9.87& True & - & 126.90 \\
       GB & Europe Gov. Bonds Fund & 10.0 & 9.97& True & - & 27.90 \\
       EQ & Global Equities Fund & 30.0 & 30.11 & True & 109.30 & -\\
       EM & Emerging Market Fund & 15.0 & 14.94 & True & - & 60.85 \\ 
       RE & European REITs ETF & 15.0 & 14.83 & False & - & 165.85 \\ 
       BT & Biotech ETF & 10.0 & 10.23 & False & 231.10 & -\\
       GD & Gold (ETC) & 10.0 & 10.04 & False & 41.10 & -\\ 
       \bottomrule
    \end{tabular}
    \caption{Example of a model portfolio (100k(€)), with transferable and non-transferable instruments}
    \label{tab:ssa}
\end{table*}

One alternative for addressing this problem is to modify the LP model to incorporate a second way for "transferring", which essentially represents a combination of sell and buy transactions that can be separated during execution. Specifically, the model is updated as follows:
\begin{itemize}
    \item include a matrix $x'_{ij}$ of duplicated decision variables to differentiate between switches and trades 
    \item  extend objective function with a synthetic cost function $c'_{ij}$ that aggregates the cost of buying and selling
    \item Update the constraints to account for moves from both types of transactions.
    \item introduce constraints $x_{k,r} = 0$ for any non-transferable instruments $k$ or $r$.
\end{itemize}

Under the additivity assumption of LP programs, this proposal is also equivalent (assuming non-equal costs) to pre-select the cheapest option between $c_{ij}$ and $c'_{ij}$ for each $i$ and $j$ and compute the solution in a single set of decision variables. Table~\ref{tab:ex_dualsol} shows a solution assuming the plausible scenario in which all possible switches are cheaper than sell+buy transactions. 

The limitation of both proposals is that they are not aware of the number of transactions. On the one hand, transaction cost would be sub-optimal if we have to include a fixed cost per transaction, as it frequently occur with brokerage fees.  On the other hand, having fewer transactions will simplify the operational burden.

One way to partially overcome this problem is to post-process the LP solution, collecting from $x'_{ij}$ all amounts in row $i$ as a single sell, and all amounts of column $j$ as a single buy.  Reviewing Table~\ref{tab:ex_dualsol}, we would have {\it (Buy 48.04 EM)} as a result of grouping BT$\rightarrow$EM and GD$\rightarrow$EM from $x'_{ij}$ assignments. 

However, this post-process adjustment can not guarantee that the LP solution is optimal in the number of transactions. Consider for instance the extreme case of having both switching and trading  costs equal for all flows.  An optimal LP solution could split the flows assigning arbitrary non-zero values in all decision variables, both $x_{ij}$ and $x'_{ij}$.  This ($i\times j\times 2$) is obviously above the optimal number of transactions and the post-process adjustment only takes care of buy and sell trades independently. Given that in many cases this proposal should provide a fairly good solutions, we will consider it for comparison in the evaluation section.

\begin{table}[]
    \centering
    \begin{tabular}{lrrrr} \hline 
    {} &     MM &    GB &     EM &      RE \\ \hline 
    EQ &  \textit{96.49} &   0.0 &  \textit{12.81} &    0.00 \\
    BT &   0.00 &  27.9 &  37.35 &  165.85 \\
    GD &  30.41 &   0.0 &  10.69 &    0.00 \\ \hline 
    \end{tabular}
    \caption{Example solution for a portfolio update. In italics 
    the move that are truly transferable}
    \label{tab:ex_dualsol}
\end{table}


\section{State Space Model}
\label{sec:state_model}
In this section we describe the proposal to formalize the multi-type transaction updates. We define the task by means of graph search in a state space. A search task is composed of the following elements:
\begin{itemize}
    \item $\mathcal{S}$ is a finite set of numeric state variables
    derived from the list L of the portfolio holdings.  
    Variables in $\mathcal{S}$ are partitioned in:
    \begin{itemize}
        \item $\{u_1,\dots,u_n$\} as the outflow variables
        \item \{$v_1,\dots,v_m$\} as the inflow variables
        \item \{$w$\} an additional special variable representing the cash balance
    \end{itemize}
    A state S in the state     space is a value assignment for all variables in $\mathcal{S}$.
    \item $\tau: L \rightarrow \{\top,\bot\}$ is boolean function representing if a given holding in L is transferable
    \item  $C_{K}: L \rightarrow \mathbb{R_+}$ is the individual trading cost as a percentage of the transaction amount. 
    \item  $C_S: L \times L \rightarrow \mathbb{R_+}$ is the combined switching cost as a percentage of the funds being transferred from one holding to another
    \item $C_F: L \rightarrow \mathbb{R_+}$ is the individual fixed cost, independent of the instrument type
    \item $I$ is the initial state, which corresponds to the initial flows 
\end{itemize}

We use the notation $u[x]$ (or $v[x]$) to indicate the value assignment of $x \in L$, when we want to refer to variables by the holding they represent.  The definition of the goal is implicit, meaning that for all tasks, the objective is to find a state $G_0$ with no pending flows. This is, a state in which $u_i=0$ and $v_j=0$ for all $i$ and $j$. 

Now we have to define the set of actions (transactions) and the associated transition function that allow us to change one state into another. 
Note that $\mathcal{S}$ entails an infinite state space. However, knowing in advance the desired values in the goal state, it does not make sense to allow actions to perform numeric changes arbitrarily.  Let us consider the SELL action scenarios.  If an outflow $x$ in any given state is not zero, the SELL $x$ action should only consider the whole amount in $u[x]$, given that if an additional SELL is needed it could have been included in the first SELL. This does not mean that the action will always refer to the same amount. As counter example, a former SWITCH can reduce $u[x]$ and then, the SELL action will still refer to the (pending) whole amount, but different from the original statement.  This simplification is also inspired from the similarities with the transportation domain. The general idea consists of forcing the model to consider only the actions that move the quantities that are either required or available.

Following this idea, the transition function that changes state S into S' when an action is applied, can be succinctly expressed as the set of parametrized operators indicated in Figure~\ref{fig:search-operators}.

\begin{figure}[hbt]
    \centering
    {\small 
\begin{tabular}{r|l}
   \toprule
   \textbf{  action} & \textbf{SELL} \\
     parameters & $< x >$ \\
     precondition & $u[x] > 0 $ \\
     effects & $w' = w + u[x]$\\
             & $u'[x] = 0$ \\ 
     cost & $C_F(x) + C_K(x)u[x]$ \\ 
    \midrule 
    \textbf{  action} & \textbf{BUY-AVAILABLE} \\
     parameters & $< x >$ \\
     precondition & $v[x] > 0 $ \\
                  & $w > 0$ \\  
                  & $v[x] - w > 0$ \\
     effects & $w' = 0 $\\
             & $v'[x] = v[x] - w$ \\ 
     cost & $C_F(x) + C_K(x)w$ \\ 
    \midrule
    \textbf{  action} & \textbf{BUY-NEEDED} \\
     parameters & $< x >$ \\
     precondition & $v[x] > 0 $ \\
                  & $w - v[x] \geq 0$ \\
     effects & $v'[x] = 0 $\\
             & $w' = w - v[x] $ \\
     cost & $C_F(x) + C_K(x)v[x]$ \\              
    \midrule
     \textbf{  action} & \textbf{SWITCH-AVAILABLE} \\
     parameters & $< x, y>$ \\
     precondition & $\tau(x) \wedge \tau(y)$ \\ 
                  &  $u[x] > 0$ \\
                  &  $v[y] > 0$ \\
                  &  $v[y] -  u[x] > 0 $ \\
     effects & $v'[y] = v[y] -  u[x] $\\
             & $u'[x] = 0$ \\ 
     cost & $C_F(x) + C_F(y) + C_S(x,y)u[x]$ \\ 
    \midrule
     \textbf{  action} & \textbf{SWITCH-NEEDED} \\
      parameters & $< x, y>$ \\
     precondition & $\tau(x) \wedge \tau(y)$ \\ 
                  &  $u[x] > 0$ \\
                  &  $v[y] > 0$ \\
                  &  $u[x] - v[y] \geq 0 $ \\
     effects & $v'[y] = 0 $\\
             & $u'[x] = u[x] - v[y]$ \\ 
     cost & $C_F(x) + C_F(y) + C_S(x,y)v[x]$ \\ 
    \bottomrule
\end{tabular}}
    \caption{Operators describing the model transition function}
    \label{fig:search-operators}
\end{figure}

A solution to this task is the update plan, the sequence of actions $\pi = (a_1, \dots, a_k)$ that transform the initial flows into the target $G_0$. Table~\ref{tab:ex_planupdate} shows an update plan for the running example.

\begin{table}[]
    \centering
    \begin{tabular}{cl} \\ \toprule 
    STEP & ACTION \\ \midrule 
    1 & SWITCH-AVAILABLE 109.3 EQ to MM \\
2 &  SELL 231.1 BT \\
3 &  SELL 41.1 GD  \\
4 &  BUY-NEEDED 165.85  to RE \\
5 &  BUY-NEEDED 60.85  to EM \\
6 & BUY-NEEDED 27.9  GB \\
7 & BUY-NEEDED 17.6  MM \\ \bottomrule
    \end{tabular}
    \caption{Action plan for updating the example portfolio}
    \label{tab:ex_planupdate}
\end{table}

\section{Search Algorithm and Heuristics}
In this section we describe the alternatives to solve a task in the search
model described before. The first point to emphasize is that all actions
have at least one effect that makes moves towards $G_0$ 
and there is no effect in the opposite direction.  Therefore, a depth-first
search (DFS) will provide a first sub-optimal solution without any single backtracking step.  Any further exploration that consider this upper-bound
should provide subsequent solutions of improving costs. Thus, the first
alternative we examine is performing a Depth-first Branch and Bound (DFBnB)
algorithm \cite{balas1983branch} until the search is exhausted or a search limit is reached (i.e., execution time or node generation count). In the case the algorithm
exhausts the search space, the last solution found corresponds to the optimal
solution. However, the state space produces a lot of symmetries, which will
cause DFBnB to scale poorly in terms of the state size.

Another option is to use a heuristic search algorithm such as A* \cite{hart1968A-star}. If we provide an admissible heuristic, the solution found by A* is optimal.  To derive such a heuristic function (Eq.\ref{eq:hfee}), we compute for any given state S, what is the lower bound cost to achieves $G_0$. Fixed fees are associated to each position, so at least one action for each of the pending non-zero variables ($L_{\rm out}(S)$ and $L_{\rm in}(S)$)
will require this fee to be paid (Eq.~\ref{eq:hfix}). For the variable costs, we consider the minimum between trading or switching funds in $C_{\rm min}$, and then, only pending outflow variables $L_{\rm out}$ are used in $h_{\rm rel}$ cost computation (Eq.~\ref{eq:hrel}) to avoid double counting in hypothetical switching transactions. At the end, the heuristic value for S is the sum of the estimated lower-bounds for both fixed and variable costs.

$$L_{\rm out}(S) = \{ x \in L \, |\, u[x] > 0 \} $$
$$L_{\rm in}(S) = \{ x \in L \mid v[x] > 0 \} $$
$$C_{\rm min}(x) = \min_{x \in L}\{C_K(x), \min_{y \in L}\{C_S(x,y)\}\} $$
\begin{equation}
\label{eq:hfix}
h_{\rm fix}(S) = \sum_{x \in L_{\rm out}(s)} C_F(x) + \sum_{x \in L_{\rm in}(s)} C_F(x) 
\end{equation}
\begin{equation}
\label{eq:hrel}
h_{\rm rel}(S) = \sum_{x \in L_{\rm pout(S)}} C_{\rm min}(x) \times u[x]     
\end{equation}
\begin{equation}
\label{eq:hfee}
h_{\rm fee}(S) = h_{\rm fix}(S) + h_{\rm rel}(S)     
\end{equation}

This heuristic function can be computed efficiently during search. $C_{\rm min}$ is not state-dependent so it is computed in advance. The rest are arithmetic computation that are linear in the state size. Extending the previous reasoning we derive an estimate of the number of transactions to achieve the goal (Eq.~\ref{eq:hcount}). Even in the best-case scenario of making all switches, at least $h_{\rm count}$ transactions are needed to complete the update.

\begin{equation}
\label{eq:hcount}
h_{\rm count}(S) = max\{|L_{\rm in}|, |L_{\rm out}| \}
\end{equation}

Nonetheless, we will not use this heuristic. We want to have the minimum number of transactions, but as a secondary objective once the optimal transaction cost is determined.  

\section{Experimental Evaluation}
In this section we present the experiments conducted to evaluate the various  solutions proposed in previous sections. The main objective is to verify whether the linear programming and search approaches produce better solutions than  the one generated by a naive approach in the multi-type transaction scenario. 
The software was implemented in Python. The LP approaches were modelled using the CVXPY library \cite{diamond2016cvxpy,agrawal2018rewriting}. The heuristic search approach was implemented as described in the the state model section. We used the SimpleAI library functionality to implement the domain transition, cost and heuristic functions. The A* algorithm is included in the library,  The DFBnB algorithm is not present, but we coded it using the library search data structures.  The SimpleAI is, as stated by the authors, a more stable pythonic implementation of the algorithms in Russell and Norvig's book \cite{russell2016artificial}. 
The list of evaluated algorithms/configurations is:
\begin{itemize}
    \item \textbf{Naive}: The simple base line of creating a plan by selling all positions with outflows and buying all positions with inflows.
    \item \textbf{LP+}: The LP model, along with the post-process adjustment for grouping market transactions, as described at the end of Section~\ref{sec:lp-dual}.
    \item \textbf{DFBnB}: Our implementation of DFBnB, providing the initial depth-limit equal to the length of the naive solution. This length is computed in advance as it is the number of flows in $I$. Additionally, the limit of generated nodes is set to 100k.
    \item \textbf{Astar-fee}: Running A* with the $h_{\rm fee}$ heuristic function (Eq~\ref{eq:hfee}).   
\end{itemize}

For all configurations, we measured the solution cost and the plan length. For search algorithms, we measured the number of generated nodes.  For DFBnB, we also measured the same metrics at the first solution found.  

First, we analyze the overall performance and the task scalability. To simulate diverse update tasks, we generated groups of random problems of incremental portfolio size, ranging from 4 up to 13 holdings. All portfolios were scaled to have 10k of money value. Each group consists of 20 problems. Each problem has the following features:
\begin{itemize}
    \item 70\% of the holdings represent transferable funds. The rest are considered a general type of ETFs.
    \item The current portfolios are allocations randomly sampled from a uniform distribution and scaled to sum 1 (i.e., no initial cash positions).   
    \item Target portfolios were designed such that random fund flows have a common factor amount.  Here we wanted the simulation to also include rebalance splits (e.g., 2\% overweight in fund $X$ is allocated 1\% to fund $Y$ and 1\% to fund $Z$.) 
    \item  ETFs have a fixed fee selected from the alternatives (0.5 1.0,  2.5) with the idea of simulating different exchange commissions. Mutual funds have no fixed fee for the subscription or redemption of shares.
    \item All funds have a variable fee in basic points (bps) ranging  from 1 to 10. In this case, we wanted to have differences between funds, to represent both explicit and implicit estimated costs. 
\end{itemize}

\begin{figure}
    \centering
    \includegraphics[width=0.9\linewidth]{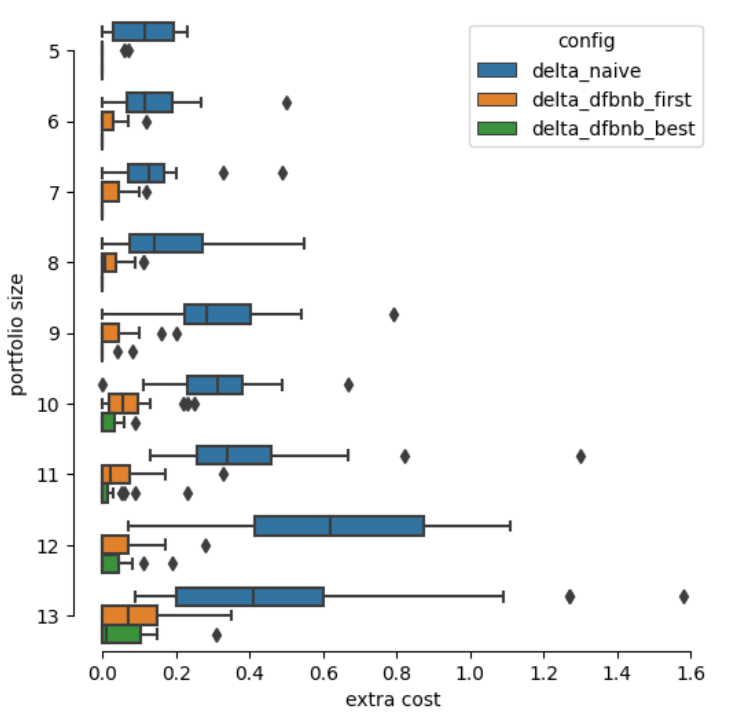}
    \caption{Distributions of costs above the the optimal cost achieved by LP+/Astar-fee}
    \label{fig:extra_cost}
\end{figure}

In 180 problems ($20\times 9$) both LP+ and Astar-fee achieved the optimal solution. As a reference, the average cost per transaction is $0.27\pm0.08$. We computed the extra costs achieved by the rest of algorithms.
Figure~\ref{fig:extra_cost} shows the distribution per portfolio size.
Naive approach is clearly producing plans of worse cost across the
whole range of portfolios. Their update plans coincide with the optimal
cost in 16 problems only.  DFBnB achieved remarkable improvements, even with
its first solution. The DFBnB best solution matched the optimal solution in 145 tasks, of which in only 74, the complete search tree was explored.  

Regarding the plan length, LP+ only matched the number of transactions with Astar-fee in 148 plans, confirming our initial conjecture that LP+ lacks awareness of the plan length. DFBnB matched the length on 174 plans and Naive in 106, but for both of them, having the same plan length as Astar-fee does not mean they achieved a solution with the same (optimal) cost.  DFBnB produced 3 fee sub-optimal plans that have fewer number of transactions.  To have a closer look of this plan length awareness we run another experiment described later in this section. 

\begin{figure}
    \centering
    \includegraphics[width=0.75\linewidth]{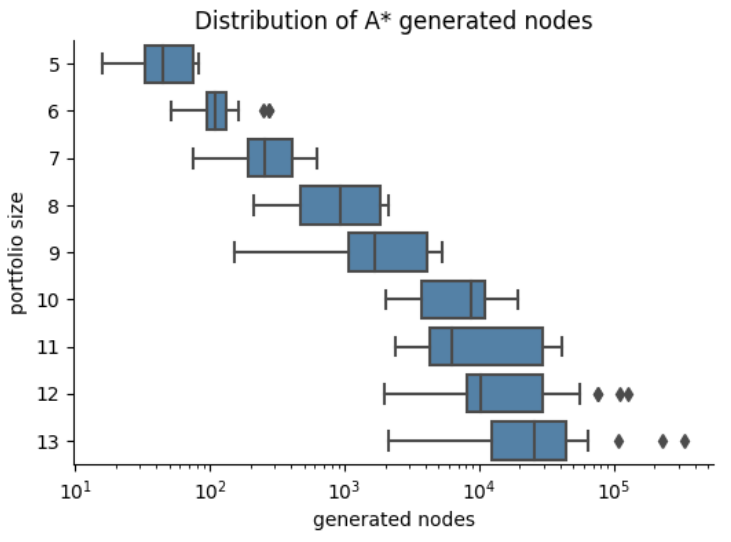}
    \caption{Distributions of generated nodes by A* on increasing portfolio size}
    \label{fig:generated-nodes}
\end{figure}

On the other hand, search algorithms have a scalability issue. They generate a number of nodes that is exponential in the portfolio size. Figure~\ref{fig:generated-nodes} shows for Astar-fee the distribution of generated nodes (in log scale) per problem size.  Nevertheless, heuristic $h_{\rm fee}$ provides relative good guidance toward the goal. Consider for example that  DFBnB explored completely all group tasks up to size 7, and 12 search trees reached the bound of 100k nodes in tasks of size 8. Besides, 
we think that testing on portfolios with up to 13 positions is fair enough for
our mutual fund scenarios \cite{lhabitant2002hedge}.  Nevertheless, real portfolio could be of larger size,
but what matters for the update plan is the effective number of positions being changed in the target portfolio.  Regarding LP+, the time for solving tasks is negligible. The number of variables and constraints are relatively small for the performance of state-of-the-art convex optimization solvers, which are the ones under the hood of the CVXPY library.

Now we focus on the update plan length. To facilitate the interpretation 
of the results we will explore problems of the same size. For this experiment we generated 500 random problems of portfolios with 10 holdings.  
Again, both LP+ and Astar-fee obtained the same solution cost in all problems.
As expected, LP+ produced solutions with sub-optimal number of
transactions. Table~\ref{tab:delta_step_factor_updates} shows the counts of the 500 plans, split by number of additional steps to the best plan length, which in all cases was achieved by Astar-fee search. As we increased the sample size,
we see more cases with several number of extra transactions. Analyzing these plans we observed that LP+ unnecessarily split switches of equivalent cost in several transactions. If a Astar-fee solution is computable in the available time resources, it will provide the benefit of being cost-optimal with the lowest number of transactions.

\begin{table}[]
    \centering
    \begin{tabular}{rrrr}
    \toprule
    \textbf{$\Delta$ steps}& \textbf{Naive} &\textbf{ LP+} & \textbf{Astar-fee} \\ 
    \midrule
    0 & 242 & 392 & 500 \\
    1 & 110 & 49 & 0 \\
    2 & 77 & 38 & 0 \\
    3 & 71 & 12 & 0 \\
    4 & 0 & 6 & 0 \\
    5 & 0 & 1 & 0 \\
    6 & 0 & 1 & 0 \\
    7 & 0 & 1 & 0 \\
    \bottomrule
    \end{tabular}
    \caption{Number of solutions containing zero or extra plan steps 
    (transactions) }
    \label{tab:delta_step_factor_updates}
\end{table}

\section{Updates as Automated Planning}
The state-space model for updating portfolios can also be approached as AI Automated Planning.  The interesting point of using automated planning is that tasks are described in a high-level standard language, such as PDDL2.1 (\textit{Planning Domain Definition Language}) \cite{fox2003pddl2} and the solutions are computed by domain-independent solvers called automated planners. In a PDDL task, the state space, actions and transition function is modelled in the PDDL domain, and the PDDL problem describes the initial state and the goals.  
In terms of representation richness, PDDL could enrich our model in the following features:

\begin{itemize}
    \item Timed Plan: Rather than having a sequence of steps, the planner will provide actions with a timestamp relative to a initial time $t_0$. It would be easy to recognize for instance that a group of transactions are independent and can be executed a the same time step.
    \item Durative actions: Apart from the cost, actions can have a duration. In our examples this will make the difference between almost instant buy and sell transactions, compared to mutual fund switches that can take some few days.
    \item Time Effects: In durative actions, effects should be temporally annotated. Thus, some effects can be considered instantaneous (eg., the cash balance is no longer available right away)
    or they can take place at the end of the action (eg. amounts in progress are settled at the end). 
\end{itemize}

Figure~\ref{fig:pddl-transfer-action} depicts the equivalent \textsc{switch-available} action modelled in PDDL2.1. 
Even though the modelling features of PDDL look appealing, the capabilities of available planners have discourage us to continue the research in this direction.  Many temporal planners have focused on  propositional concurrency, therefore they are only supporting a fragment of PDDL2.1 without numeric state variables. From the remaining list of available planners we found that the {\sc Optic} planner\cite{benton2012temporal} is the most reliable one in terms of handling the language features.    

\begin{figure}
    \centering
       
    {\tiny
    \begin{verbatim}
    
  (:durative-action switch_available
   :parameters (?from - outfund ?to - infund)
   :duration (= ?duration (transfer_time ?from ?to))
   :condition (and
       (at start (transferable ?from))
       (at start (transferable ?to))
       (at start (< (+ (delta_target ?from) (delta_target ?to)) 0))
       (at start (> (delta_target ?from) 0))
       (at start (< (delta_target ?to) 0))
       (at end (> (in_progress ?from ?to) 0))                          
    )
    :effect (and 
        (at start (assign (delta_target ?from) 0))
        (at start (assign (in_progress ?from ?to) 
                          (delta_target ?from)))
        (at end (increase (delta_target ?to) 
                          (in_progress ?from ?to)))
        (at end (assign (in_progress ?from ?to) 0))
        (at end (increase (total-cost)
                          (+ (* (transac_fee ?from)
                                (in_progress ?from ?to))
                          (* (transac_fee ?to)
                             (in_progress ?from ?to)))))
        )
    )
    \end{verbatim}}
        
    \caption{An PDDL action for switching transferable funds}
    \label{fig:pddl-transfer-action}
\end{figure}

We developed a complete PDDL model for portfolio update tasks.
It is equivalent to our domain-dependent implementation, but including durative actions.  However, Optic did not provide competitive solutions even for small-sized tasks. A timed plan is not justified if it implies significantly sub-optimal results compared to the domain-dependent alternative. Optic runs a sup-optimal algorithm (WA*) with a heuristic derived in a domain-independent way.   At this point
we do not think that additional effort in this direction would benefit the performance terms of the application perspective 

\section{Discussion}
\begin{figure}
    \centering
    \includegraphics[width=0.75\linewidth]{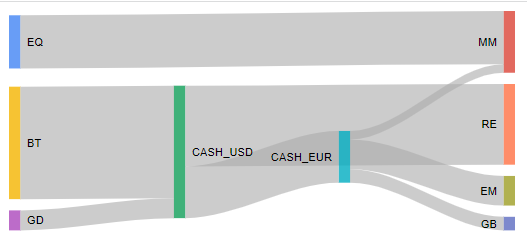}
    \caption{Graphic flow representation of the portfolio update with forex transaction}
    \label{fig:sankey_flow_fx}
\end{figure}

Now, we want to delve into some aspects that in terms of portfolio operations are somewhat closer to the real world. Our model can be extended or adapted to cover these features.  
Let's consider that in the original example the exchange-traded instruments are denominated in US Dollars (USD) and the mutual fund shares are denominated in Euros (EUR). The fund flows represent the equivalence in the portfolio base currency (EUR). Sell actions will produce USD cash balance. Therefore, to have an actionable rest of the plan, we should include a forex transaction that exchanges part of the USD for EUR, to buy shares of EUR mutual funds. 
Figure~\ref{fig:sankey_flow_fx} depicts the fund flows including the currency  exchange. The basic option is to exchange currencies as needed on a per-transaction basis.  Another option is to compute the currency imbalances from the original flows, and extend our model to include:
\begin{itemize}
    \item variables $\{ w_1, \dots , w_n\}$ instead of the single $\{w\}$, to keep record of the cash balances by currency.
    \item Operators for forex transactions. {\sc Exchange-available} to fully exchange a cash balance, and {\sc Exchange-needed} to fulfill a currency imbalance. 
\end{itemize}

Mutual funds charge a management fee that is passed on over time to the fund net asset value. This implies that in practice, few funds have a explicit commission 
at the time buying or redeeming shares \cite{greene2007daily}, and therefore in fund switches. What is left, it is building a cost function from implicit cost estimates, but this can be a difficult task without information from fund sponsors.  Although the cost function is a rough approximation, the solutions are always useful because they provide a feasible assignment of how to distribute the flows.
Any form of rebalance automation needs an update plan no matter it is optimized with accurate input costs or not.

For the particular case of the Traspasos regime, there is an alternative cost function that does not focus on the monetary cost. It turns out that the regulation allows the transfer, especially if it is between different managers, to be carried out in up to 8 days.  This means that the effective day of selling and buying the shares is not the same. Therefore, from the point of view of an investor who wants to be always invested, the cost function can be the money-weighted sum of the time out-of-the-market. However, this function is useful for the basic LP model when only switches are considered. Incorporating it into the multi-transaction version is more complicated because the almost instantaneous execution of market orders can distort its calculation.

On the other hand, we also consider the cost for the investor regarding his tax bill. Unlike the previous cost functions, this one is specific to each portfolio.  If the information is available, one must compute the capital gains implied by the initial outflows. The great difficulty is that 
these data, as it is of personal nature, will not be available for research purposes. Consequently, it becomes necessary to create high-quality simulated data to evaluate the potential impact of using an optimized solution with this type of function. Combining the cost functions into a single weighted function would allow modeling a multi-objective approach, but the mechanism for choosing the weights and the implications for the solution plans is an open question.  

Portfolio updates also poses additional challenges when integrated management has to allocate money across multiple investor accounts.  It is typical for an investor to have different accounts (tax-exempt, taxable-accounts, retirement plans). In this context, rebalancing the portfolio may imply that you first have to transfer money from one account to another, and then
carry out operations related to investment funds.  In the same way, our model can include these transfer actions between accounts, with their associated costs, if any.

Finally, the general case of multi-transaction updates could also be modeled with MIP (Mixed Integer Programming).  
With boolean variables it is possible to model for instance if an action is executed or not. The objective function could have a factor that takes into account the number of transactions. We preferred the state-space model since the definition of the actions facilitates plan interpretation, or from other view, they have a closer similarity to the transactions that are then executed. 

\section{Conclusion}
We presented a comprehensive analysis of various approaches for generating a plan to update a portfolio from its current allocation to a target allocation. We successfully developed a state-space model capable of handling transactions involving two types of funds. In particular, this model is useful to provide plans to update portfolios in which the transfer must be considered as a special transaction within the Spanish tax regime.  We showed that using heuristic search with our model allows us to generate plans that, maintaining the optimal solution, manage to have fewer transactions than a simplistic solution or an LP solution that does not take into account the number of transactions. 

Moreover, we discussed the potential extensions of our model to address specific portfolio requirements, including dealing with various currencies, execution times, and the tax implications faced by clients. As portfolio updates occur frequently, we strongly believe that optimizing back-office processes can lead to cost improvements, ultimately benefiting both clients and financial institutions.
In conclusion, our research contributes valuable insights and practical solutions to optimize portfolio management processes, fostering better performance in the fund industry.

\section{Acknowledgements}
This paper was prepared for informational purposes by the Artificial Intelligence Research group of JPMorgan Chase \& Co and its affiliates (“J.P. Morgan”) and is not a product of the Research Department of J.P. Morgan.  J.P. Morgan makes no representation and warranty whatsoever and disclaims all liability, for the completeness, accuracy or reliability of the information contained herein.  This document is not intended as investment research or investment advice, or a recommendation, offer or solicitation for the purchase or sale of any security, financial instrument, financial product or service, or to be used in any way for evaluating the merits of participating in any transaction, and shall not constitute a solicitation under any jurisdiction or to any person, if such solicitation under such jurisdiction or to such person would be unlawful.   

\bibliographystyle{aaai}
\bibliography{rebalance}

\end{document}